\documentclass[useAMS,usenatbib]{mn2e}
\usepackage{times}
\usepackage{url}
\usepackage{graphicx}
\usepackage{amssymb,amsmath}
\graphicspath{{./figs/}}

\newcommand\rg{\ensuremath{r_\mathrm{g}}}

\newcommand\lsph{\ensuremath{L_\mathrm{iso}}}

\title[Modeling the light curves of ULXs as precession]{Modeling the light curves of
  ultraluminous X-ray sources as precession}

\author[T. Dauser et al.]{T.\ Dauser$^{1}$\thanks{E-mail:
    thomas.dauser@sternwarte.uni-erlangen.de}, M.\ Middleton$^{2,3}$,
  J. Wilms$^{1}$ \\ $^{1}$ Dr.\ Karl Remeis-Observatory and
  Erlangen Centre for Astroparticle Physics, Sternwartstr.~7, 96049
  Bamberg, Germany \\ $^{2}$ Institute of Astronomy, Madingley Road,
  Cambridge CB3 0HA \\ $^{3}$ School of Physics \& Astronomy, University of Southampton, Southampton, UK } 
\begin{document}

\pagerange{\pageref{firstpage}--\pageref{lastpage}} \pubyear{2016}

\maketitle
\label{firstpage}

\begin{abstract}
  We present a freely available {\sc xspec} model for the modulations
  seen in the long-term light curves of multiple ultraluminous X-ray
  sources (ULXs). By incorporating the physics of multiple electron
  scatterings (ray traced with a Monte-Carlo routine), we go beyond
  analytical predictions and show that the geometrical beaming of
  radiation in the conical outflow can be more than a factor of 100
  for opening angles smaller than $10^\circ$. We apply our new model
  to the long-term, well sampled \textsl{Swift} light curve of the
  recently confirmed ULX pulsar \mbox{NGC~5907 X-1} with an
  established period of 78 days. Our results suggest that geometrical
  beaming together with a slight precession of the conical wind can
  describe the light curve with a consistent set of parameters for the
  wind. The small opening angle of roughly $10\mathrm{-}13^\circ$
  implies a highly super-critical flow and boosting factors at the
  order of $\mathcal{B}=60$--$90$ that would yield a fairly low
  surface magnetic field strength of $2\times 10^{10}\,$Gauss.
\end{abstract}


% ================= %
\section{Introduction}
\label{sec:introduction}
% ================= %

Ultraluminous X-ray sources (ULXs) have traditionally been defined by
both their apparent X-ray luminosities exceeding $1\times
10^{39}\,\mathrm{erg}/\mathrm{s}$ and by their spectra showing a
`break' between 2 and 10\,keV
\citep[see][]{Stobbart2006a,Gladstone2009a,Bachetti2013a}. More
recently, their short (typically $< 10$\,ksec) and long (months-years)
timescale variability has been used as a complementary lever arm to
diagnose the nature of the accretion flow with spectral-timing results
suggesting that the majority of the bright
($>3\times10^{39}\,\mathrm{erg}/\mathrm{s}$) ULXs can be successfully
interpreted in the framework of super-critical accretion
\citep[see][]{Shakura1973a,Poutanen2007a} onto stellar remnants
($<100M_{\odot}$) \citep[see][and references
therein]{Sutton2013a,Middleton2015b}.
 
Numerical radiative magnetohydrodynamical (RMHD) studies of
super-critical accretion
\citep[see][]{Ohsuga2007a,Ohsuga2011a,Jiang2013a,Sadowski2014a}
confirm the analytical prediction that the scale-height of the
accretion flow is very large (typically $\sim$1) when the mass
transfer rate through the disc (typically
$\gtrsim10^{-6}\,M/10\,M_{\odot}$) results from unstable mass transfer
occurring when the donor star evolves on a thermal timescale
\citep{King1999a,King1999b,King2000a,Podsiadlowski2000a,Tauris2000a}. The
large scale-height disc and subsequently radiation pressure driven
winds are expected to precess on long timescales (see Middleton et
al., in prep).  The predicted spectral-timing signature of such
precession (as well as changes in accretion rate) was formalised in
\citet{Middleton2015b}. Such predictions are consistent with the
observed spectral-timing evolution seen in canonical systems
(e.g. \mbox{HoIX~ULX-1} \citep{Luangtip2016a}, whilst additional
evidence for precession comes from the correlation between the
strength of atomic features --- indicating an outflowing, mass loaded
wind \citep{Middleton2014a,Pinto2016a,Walton2016a} --- and the
spectral hardness \citep{Middleton2015c}.
 
A corollary of such a large scale-height flow is that the emission
from the inner regions is trapped and geometrically beamed out of
the wind-cone \citep{King2009a}, which, depending on the opening angle
should lead to orders of magnitude amplification in observed
brightness. Such beaming must follow from the presence of an
equatorial optically thick wind \citep{Pinto2016a} and is consistent with the general lack of eclipses in ULXs \citep[although see the
  recently discovered eclipsing ULXs by
  \citealp{Urquhart2016a}]{Middleton2016a}.

Numerical simulations to-date have focused on how the radiation is
generated in the flow (from dissipation due to magnetorotationally
induced turbulence) and the impact of vertical as well as radial
advection \citep{Jiang2013a} but \emph{not} the impact of geometrical
beaming as a function of the wind opening angle. In this \emph{paper}
we determine the observational impact of scattering off an optically
thick wind cone and the shape such precession should leave in the long
timescale light curves of such sources
(Sect.~\ref{sec:simulation}). In Sect.~\ref{sec:model} we present an
{\sc xspec} local model built from our simulations that can be used to
model the light curve variations. We then apply this model
successfully to data of the ULX pulsar \mbox{NGC~5907 ULX-1} in
Sect.~\ref{sec:discussion}, before providing conclusions and a summary
in Sect.~\ref{sec:summary-conclusions}.

% ================= %
\section{Simulations} 
\label{sec:simulation}
% ================= %

While radiation MHD simulations show the outflowing winds in ULXs to
be fairly complex on small scales \citep[e.g.,
][]{Takeuchi2013a,Jiang2014a,Sadowski2014a}, they confirm the global
geometry to be conical. In order to determine the impact of beaming
for a given wind opening angle, we therefore employ a simple, but
representative toy model. We assume that due to the extremely large
accretion rate, the outflow lying above the radiation pressure
supported accretion disc forms a cone and is fully ionised down to a
Thomson depth greater than unity. Radiation emitted from the inner
regions of the cone will be efficiently scattered off its surface. We
do not consider the spectral changes (moderate Compton
down-scattering) and so treat the geometrical beaming as
achromatic. The scattering itself is modelled by the Klein-Nishina
formula in a razor-thin layer of the conical surface, as will be
described later in detail.

\begin{figure}
  \centering
  \includegraphics[width=\columnwidth]{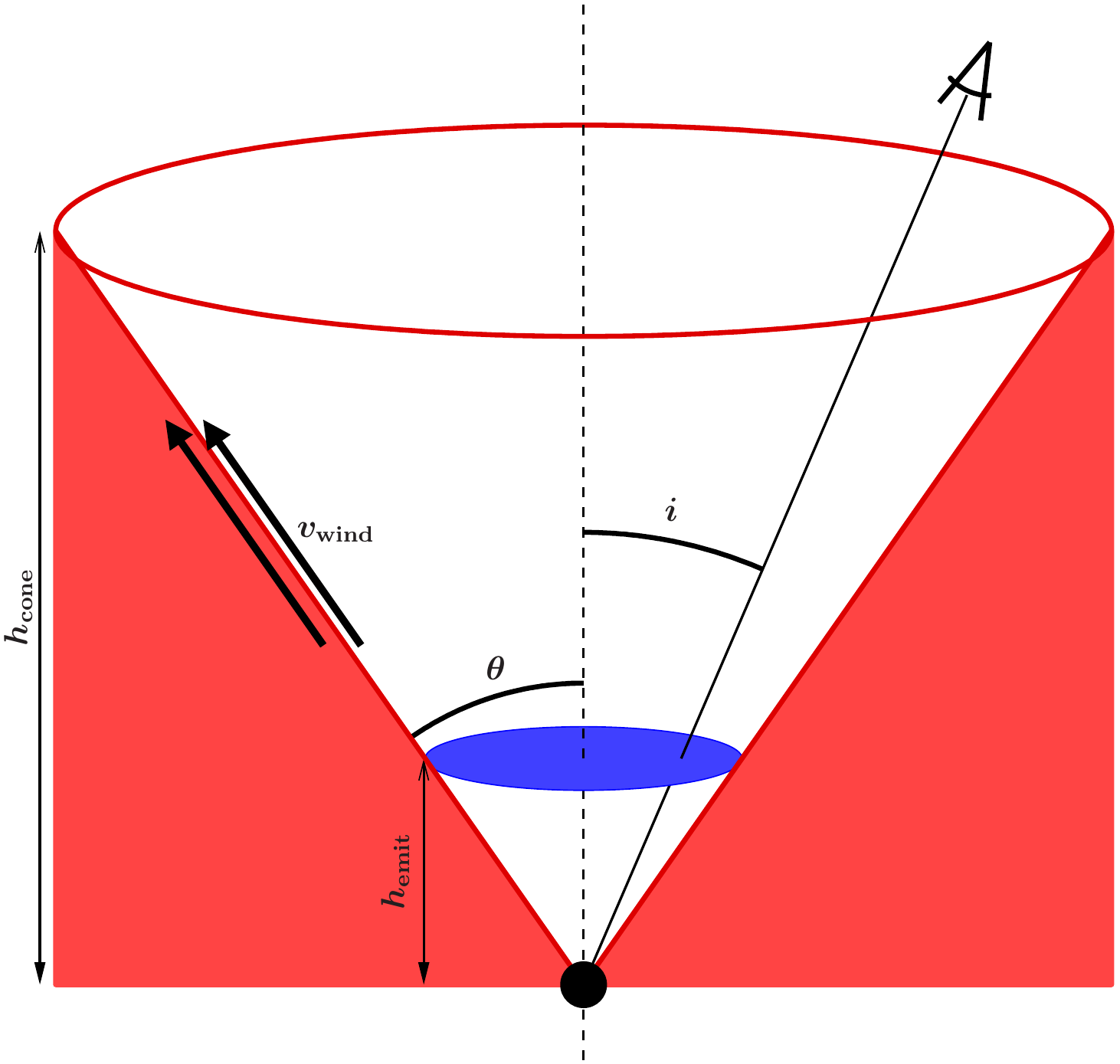}
  \caption{Geometrical model of the super-critical accretion disc 
    (drawn in red), labelling the important parameters of the model,
    the cone's half opening angle, $\theta$, the height of the cone,
    $h_\mathrm{cone}$, and the observer's inclination angle, $i$. The
    emitting region is shown as the blue region, which is located at a
    height $h_\mathrm{emit}$ above the black hole. The velocity of the
    wind at the surface of the cone is given by $v_\mathrm{wind}$.}
  \label{fig:setup}
\end{figure}

A sketch of the model geometry, including the definition of the
parameters used to describe it, is depicted in Fig.~\ref{fig:setup}.
The geometry is very much in agreement with the model proposed by
\citet[Fig.~1]{Middleton2015b}, focusing only on the radiation emitted
in the cone and the scattering off its surface. We parameterize the
cone by the angle, $\theta$, between the rotational axis and the
surface of the outflow. The opening angle of the cone is therefore
$2\theta$. The height of the cone is described by the parameter
$h_\mathrm{cone}$ which according to standard theory is related to the
physical properties of the accretion flow by $h_\mathrm{cone} \sim
\dot{m} r_\mathrm{in}$, where $\dot{m}$ is the mass inflow rate
through the disc, and $r_\mathrm{in}$ is the inner disc radius
(approximately the ISCO). Note that this relation is only valid for
the inflow (where the scale-height is radiation pressure supported and
approximately unity) and does not necessarily hold for the direction
and height of the outflow.

We assume the emission to originate isotropically from a disc-like
region located $h_\mathrm{emit}$ above the central object (as shown in
Figure 1), whilst $h_\mathrm{cone} \gg h_\mathrm{emit}$. This is an
approximation to emission from the region where the outflow is
optically thin \citep[see][]{Poutanen2007a}. As we will see, the
resulting light curve is insensitive to the exact shape of the
emission region.  The emission itself is isotropic in all directions,
i.e., also towards the black hole. The simulation and calculations for
the model are performed using Monte Carlo simulations in three
dimensions. The observer's inclination to the system is defined to be
the angle between the rotational axis of the black hole and the
line-of-sight.

We locate the emitter at a height of $10\,r_\mathrm{g}$ to be of order
the radial location of the inner photosphere $R_\mathrm{ph,in}$
\citep[see][]{Poutanen2007a}, where the outflow is starting to become
optically thick and form the surface of the cone in our model. This
has the benefit of being large enough such that general relativistic
effects are secondary to the scattering processes taking place in the
cone and can be safely ignored. The photons are therefore assumed to
travel on straight, Euclidean, trajectories between their interactions
with the cone.

The physics of the surface of the accretion disc, especially in such
an inflated state is currently not well constrained. As discussed in
the introduction, due to the large quantity of photons trapped inside
the cone it is expected that the surface is highly ionised
\citep[values expected up to $\xi > 10^6$, see, e.g.,
][]{Middleton2015b}. We therefore assume that the scattering on the
surface of the cone is purely due to electron scattering, described
using the Klein-Nishina differential cross section. While each
encounter of a photon with the surface of the cone would actually lead
to multiple scatterings inside the surface layer, in the current
approach this is simplified to a single scattering event on a razor
thin surface. This simplification can be made, as we are only
interested in the direction of the photon and therefore the
Klein-Nishina angle characteristic of the scattering will still be
conserved. Note that choosing an isotropic scattering profile instead,
does not largely influence the simulation results.  A full treatment
of the scattering including the energies is out of scope for this
\emph{paper} and will be subject of future investigations.

The simulation also includes special relativistic boosting of the
photons, in order to model an outflowing surface of the cone
\citep[see][]{Poutanen2007a}. It is taken into account by transforming
the direction of a given photon from the rest frame of the cone, where
the scattering is simulated, to the global, non-moving frame of the
system. Rough observational constraints (lower limits) on the wind
speed lie between \mbox{$v\approx0.2$--$0.4\,c$} \citep[see,
  e.g.,][]{Middleton2014a, Pinto2016a}.

We note that emission from the outflow itself (i.e. scattered emission
from the underlying thick disc) is ignored. This is a reasonable
first order assumption as the inner regions are those from which
emission is most likely to be beamed.

% ================= %
\section{Results}
\label{sec:results}
% ================= %

\begin{figure*}
  \centering
  \includegraphics[width=\textwidth]{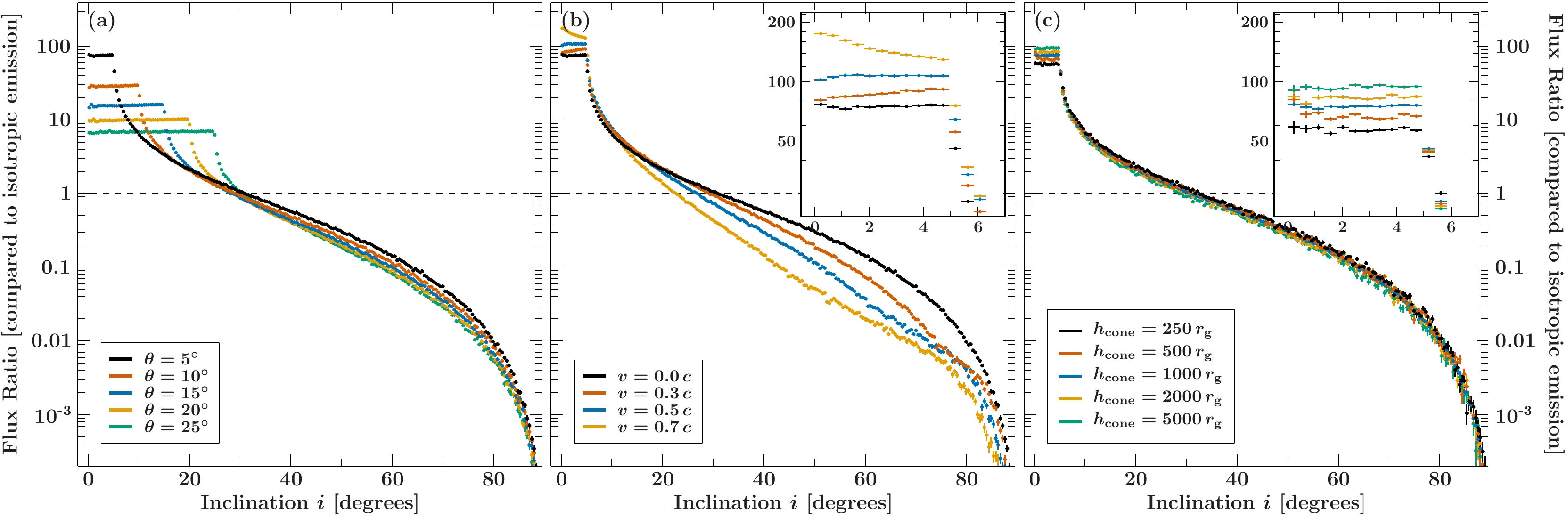}
  \caption{Flux emerging from the outflow compared to an isotropically
    emitting source, shown for different inclination angles (a)
    Variation of the of the flux amplification as a function of the
    half-opening angle, $\theta$. The height of the cone is chosen at
    an intermediate value of $h_\mathrm{cone}=1000\rg$. (b) Variation
    of the flux for different velocities of the cone, assuming fixed
    height of the cone of $h_\mathrm{cone} = 1000\,\rg$ and
    $\theta=5^\circ$. The inset shows a zoom into the low inclination
    angle region. (c) Emissivity profile for a non-moving surface
    ($v=0$) for increasing height $h_\mathrm{cone}$ of the cone, and
    again $\theta=5^\circ$.}
  \label{fig:iso_angle_param}
\end{figure*}

The simulation set-up described above directly predicts the viewing
angle-dependent flux from a specific system. The results for
different parameter combinations are shown in
Fig.~\ref{fig:iso_angle_param}. As mentioned in the previous section,
velocities of the outflow are expected to be in the range of
\mbox{$v\approx0.2$--$0.4\,c$} and we consider a height of the cone of
1000\,\rg\ as plausible. Note that even much larger values have been
suggested for SS433 \citep{Fabrika2011a}. Currently there
do not exist any constraints on the opening angle of such a cone.

The number of scatterings a single photon exhibits until escaping the
cone strongly depends on the opening angle of the cone.  In cones with
small opening angles of $10^\circ$ photons will usually scatter around
$\sim50$ times. This number quickly drops to only around $\sim 15$
scatterings for $20^\circ$, and 1 scattering on average for
$45^\circ$. The exact number of scatterings depends on the height of
the cone and also the velocity of the outflow.

We determine the effect of the boosting in the wind-cone by normalising
the outgoing radiation relative to the flux of an isotropic source,
defining the \emph{flux boost} as
\begin{equation}
  \mathcal{B} = \lsph(i) / L \quad,
\end{equation}
where $L$ is the actual luminosity of the source, and $\lsph(i)$ the
luminosity measured at a certain inclination $i$ to the system
assuming an isotropically emitting source.  The latter value is
generally quoted as the observed luminosity for ULXs due to the lack
of geometrical information. Therefore $\mathcal{B}$ actually describes
the amount by which the luminosity would be overestimated\footnote{It
  is therefore the inverse of the beaming factor $b$ used by
  \citet{King2009a}.}

Figure~\ref{fig:iso_angle_param}a shows the flux boost for a
{\it stationary} cone and different opening angles. It is immediately
evident that for $i\le\theta$, i.e., when we directly look into the
cone, the observed flux is constant, but boosted strongly depending on
the opening angle. For values of $\theta$ as small as $5^\circ$, an
effective flux boost of up to a factor $\approx$100 is possible simply
by collimation. If the observer is not
looking directly into the cone, the observed flux drops quickly below
the value expected for isotropic emission. For larger observer
inclination angles (i.e. $i\ge\theta$) the emissivity profile does not
depend on the opening angle.

Besides the system inclination, the emissivity profile also depends on
the geometrical and physical properties of the cone. The two
parameters mainly influencing the profile are the velocity of the
outflow (Fig.~\ref{fig:iso_angle_param}b) and the vertical extent of
the cone (Fig.~\ref{fig:iso_angle_param}c). Generally the dependence
of the boost on {\it both} parameters is not very strong, although a
large parameter space is explored in the plots. Notably, a change in
the height of the cone will be strongly degenerate with the
normalisation of the curve and therefore introduce an uncertainty in
the determination of the boosting factor. However, the inclination
angle at which the steep drop in emissivity occurs is unaffected. The
flux boost when looking directly into the cone can be large and of the
order of a factor $\approx$2 for velocities up to
$v=0.7\,c$. Differences between the outflowing velocities at larger
inclination angles are very small and hard to detect with current
detector sensitivities. We also note that even a slightly different
shape of the cone or variability might also produce differences of
similarly small magnitudes.

\begin{figure}
  \centering
  \includegraphics[width=\columnwidth]{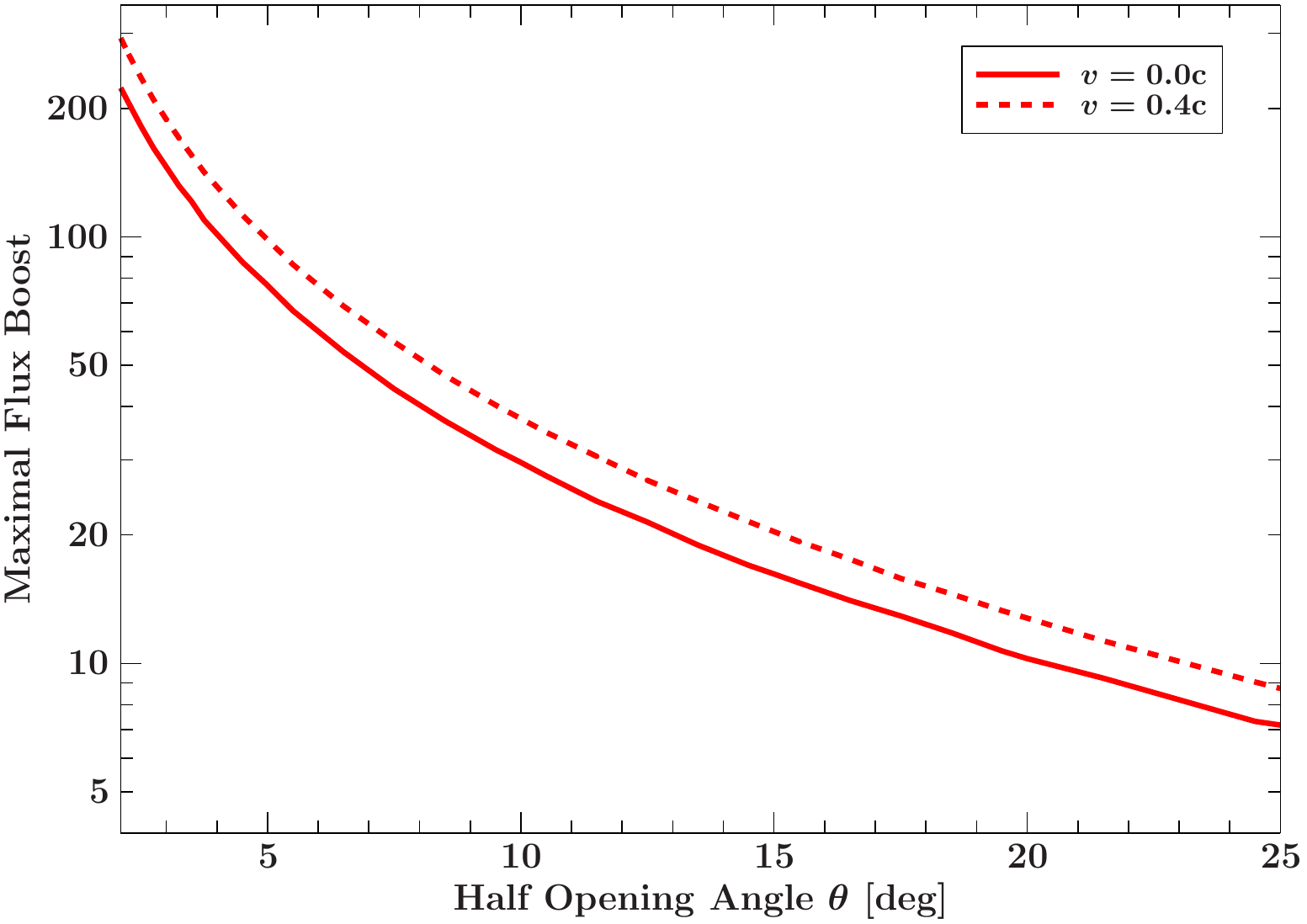}
  \caption{This plot visualises the maximal flux boost possible for a
    certain values of $\theta$ (half opening angle). Different
    combinations for a stationary (solid) and fast moving surface
    (dashed) of the cone are shown. }
  \label{fig:fluxboost}
\end{figure}
From the previous figures we can conclude that there is a flat part of
the emissivity profile when we look directly into the cone, where, due
to scattering on the outflowing walls of the cone, the intensity is
geometrically beamed. Assuming that we observe a source when looking
into the cone, the apparent maximal flux boost
$\mathcal{B}_\mathrm{max}$ for a certain configuration can be
calculated. Figure~\ref{fig:fluxboost} shows the flux boost for a
stationary and outflowing wind at a height of 1000\,\rg. While for
very narrow cones a factor of a few hundred in flux boost is possible,
even for an opening angle of $30^\circ$ (i.e., $\theta=15^\circ$) a
source would be observed at a factor of $\approx 10$ boosted compared
to its isotropic luminosity.

% ================= %
\section{Precessing discs in ULX}
\label{sec:model}
% ================= %

Observations of ULXs have revealed roughly periodic flux variations on
the order of months in sources such as M82~X-1, \citep{Kaaret2006a},
M82~X-2 \citep{Kong2016a} and NGC~5907~ULX-1 \citep{Walton2016a}. The
latter source has very recently been identified as a pulsar accreting
at a highly super-Eddington rate
\citep{Israel2016a,Fuerst2016a}. Using the above model of geometrical
beaming in the cone of the outflow, such a behaviour can be readily
explained, either by changes in the line-of-sight to the observer due
to precession of the binary orbit, or (as suggested in the case of M82
X-1 \citet{Pasham2013a}), more likely by a precession of the accretion
disc. The existence of precessing discs has long been known
\citep[e.g. in Her X-1, see][]{Petterson1977a} and also investigated
from a theoretical point of view
\citep[see][]{Pringle1996a,Maloney1997a,Maloney1998a,Fragile2007a}. Recent
studies beyond the basic theory suggest that also self-gravity of the
disc influences the precession in the case of supermassive black holes
\citep{Tremaine2014a}. In addition, the low-frequency quasi-periodic
oscillations (QPOs) found in binaries have been recently connected to
Lense-Thirring precession of the accretion flow
\citep[see][]{Ingram2009a,Ingram2011a}.

We now study whether we can reproduce similar light curves using a
precessing inflow/outflow, where the inclination with respect to the
accretion disc's angular momentum vector (or the system orbit) is
varying. We will designate this
variation by $\Delta i$, i.e., effectively the system is viewed from
angles in the range from $i - \Delta i$ to $i + \Delta i$.

Using the approach outlined above, we constructed an {\sc xspec} local
model \citep{Arnaud1996a}.  The model is publicly available for
download\footnote{http://www.sternwarte.uni-erlangen.de/research/ulxlc}
and can be readily used in any common X-ray data analysis software.
The emissivity profiles are precalculated in a table on a fine grid,
which allows for a fast evaluation. The height of the accretion disc
is fixed to $h_\mathrm{cone}=1000\,\rg$, as for reasonable heights the
parameter is largely degenerate with changes in the flux normalisation
and does not have a large effect on the overall emissivity profile
(see Fig.~\ref{fig:iso_angle_param}).

\begin{figure}
  \centering
  \includegraphics[width=\columnwidth]{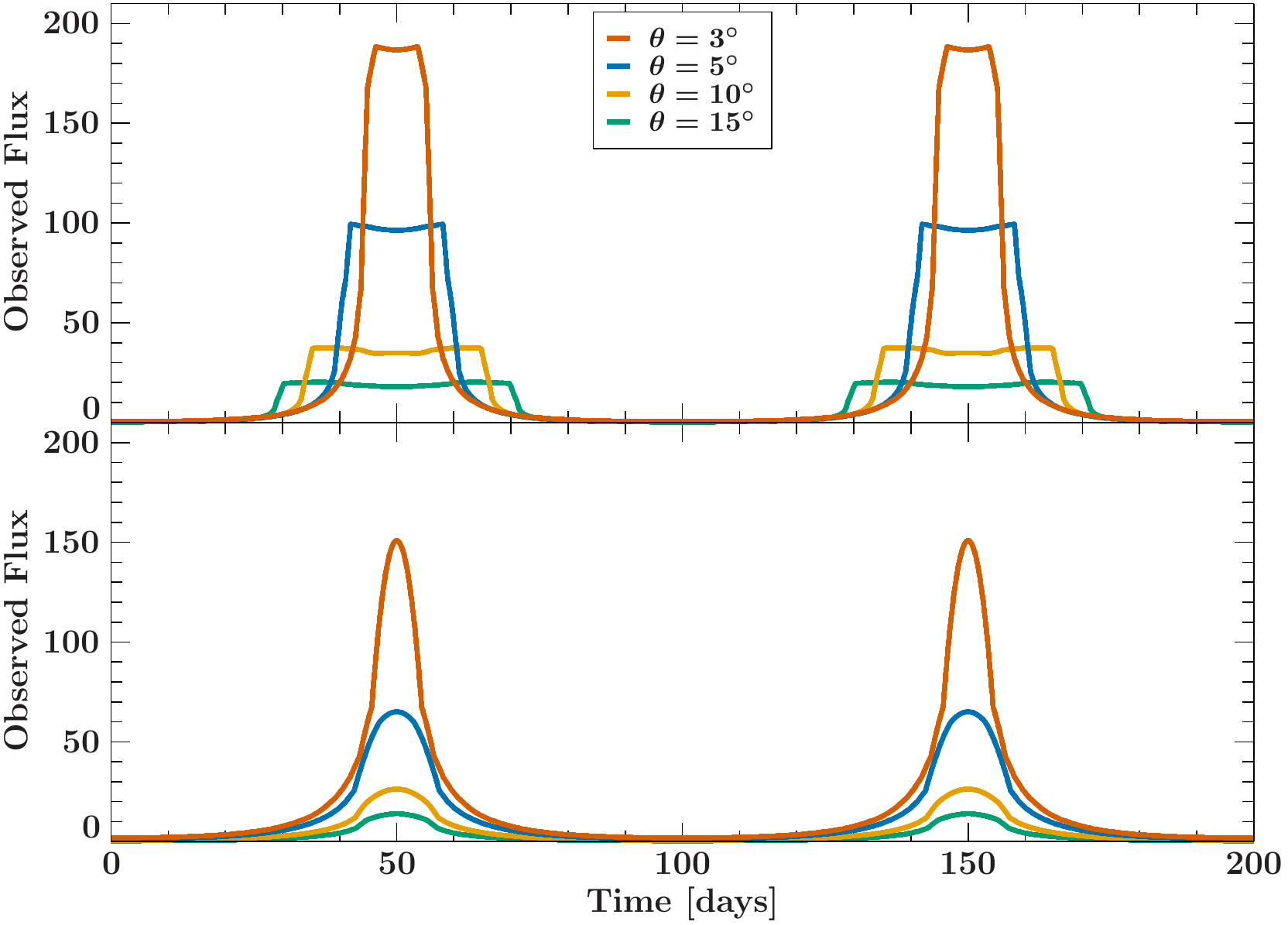}
  \caption{Simulated light curves with the \texttt{ulxlc} model, for a
    velocity of 0.4~$c$, fixing the height of the cone to
    $h_\mathrm{cone}=1000\,\rg$. The orbit is arbitrarily assumed to
    be 100\,days and we show the light curve for two precession
    periods for clarity. Different colours are for different half
    opening angles, $\theta$, as indicated in the plot. \emph{Upper
      panel:} Light curve for a larger viewing angle ($i=20^\circ$)
    and larger precession angle amplitude $\Delta i =
    18^\circ$. \emph{Lower panel:} Varying inclination angle according
    to $i = \theta + 10^\circ$ with a fixed precession angle $\Delta i
    = 10^\circ$. }
  \label{fig:lc}
\end{figure}
An example of the model in action for certain parameter combinations
is shown in Fig~\ref{fig:lc}. As expected, the duration of the peaks
increases with opening angle whilst the strength decreases. For
extreme precession angles, a double peaked profile is in principle
possible (as the observers line-of-sight passes fully over the face of
the cone).

Note that the light curve predicted in this model assumes a constant
mass accretion rate, $\dot{m}$. If $\dot{m}$ is varying, an additional
flux variation is expected (Poutanen et al. 2007). In order to model
light curves, we will assume that changes in $\dot{m}$ are small
enough to be ignored (or that variations at the spherisation radius
are long compared to the precession) such that the {\it intrinsic}
luminosity is constant over a precession period.

% ================= %
\section{Application to data}
\label{sec:discussion}
% ================= %

From the predicted light curves (see Sect.~\ref{sec:model}), it
is clear that small opening angles of around $10^\circ$ are
necessary in order to explain the boosting from a moderate luminosity
of about the Eddington luminosity, $L_\mathrm{Edd}$, to an observed
luminosity of larger than $100\,L_\mathrm{Edd}$.

In the following we apply our model to NGC~5907 ULX-1 \citep[recently
  confirmed as a pulsar][]{Israel2016a}. As revealed by high-cadence
{\it Swift} monitoring observations spanning several years, the source
shows a remarkable pattern of variability, with repeated brightenings
on timescales of $\sim$78 days \citep{Walton2016a} and {\it observed}
luminosities of over $100 L_\mathrm{Edd}$ (for a canonical neutron star
mass). We obtain the data from the UK Swift science data centre
(UKSSDC \footnotemark\footnotetext{http://www.swift.ac.uk}) for the XRT in
photon counting mode and rebin the resulting light curves to 4
days/bin. We convert the light curve into pha format using the {\sc
  ftool} {\sc flx2xsp} which also creates a flat response file.

\begin{figure}
  \centering
  \includegraphics[width=\columnwidth]{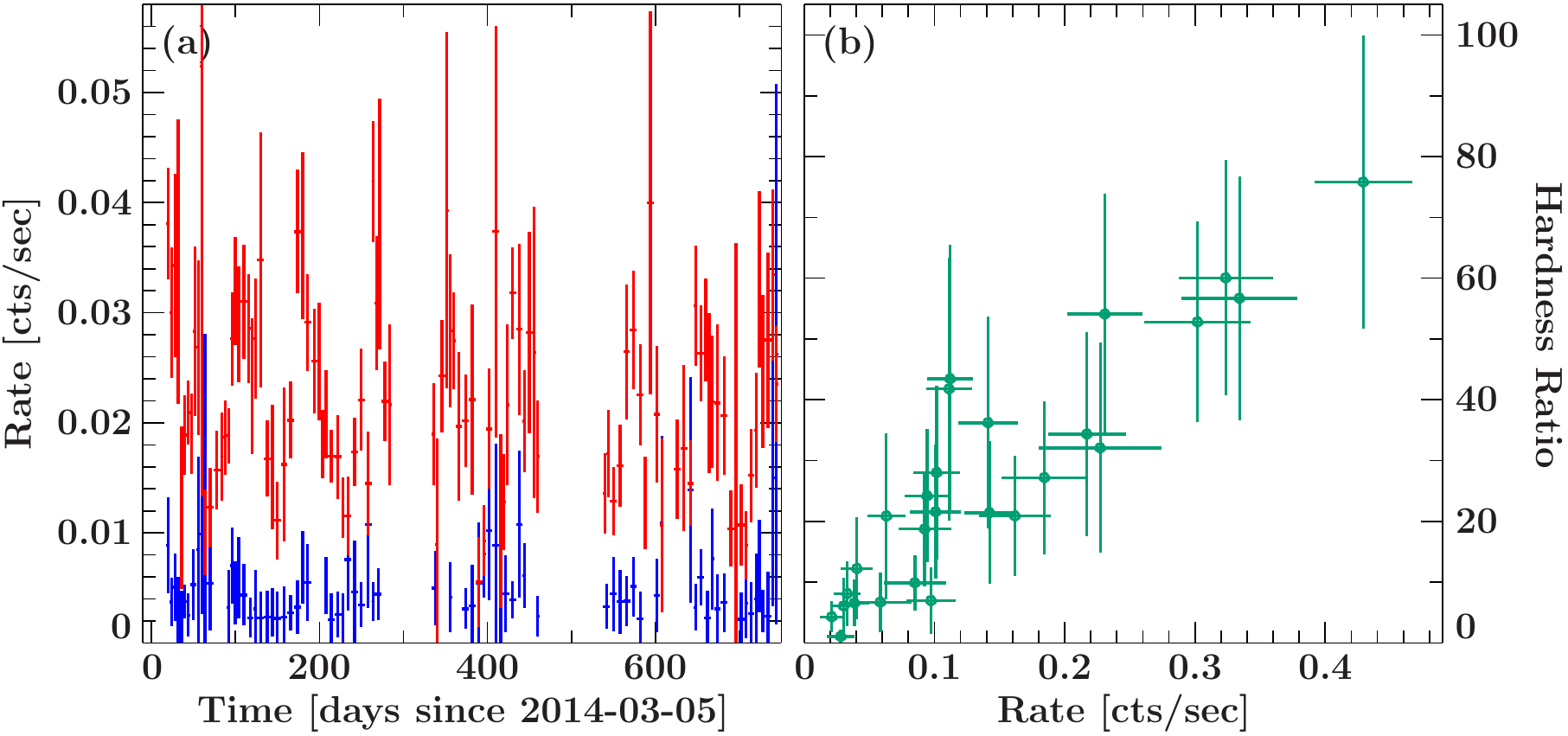}
  \caption{ \emph{Left:} long-term \textsl{Swift} light curves of the
    hard band (1--10\,keV, red) and soft (0.3--1\,keV, blue) band of
    NGC~5907 ULX-1. The binning is 4 days. It can be seen that the
    hard light curve, which shows the quasi-periodic peaks, dominates
    the total observed flux of the source. \emph{Right:} hardness
    ratio of the \textsl{Swift} count rate in the 1-10\,keV band
    divided by the 0.3-1\,keV band plotted against the total count
    rate. To increase the statistics, a binning of 25 days was chosen
    (close to 1/3 of the precession period of this pulsar). A clear
    hardening for larger count rates is evident. }
  \label{fig:hardSoft}
\end{figure}
First we obtain the light curve in soft (0.3-1\,keV) and hard
(1-10\,keV) energy bands in order to verify that the hard X-rays coming from the
innermost accretion flow indeed produces the long-term
variations. Figure~\ref{fig:hardSoft}a show both light curves where it can clearly be seen that the quasi-periodic peaks
are only found in the hard (red) light curve, which is also largely
dominating the total observed flux. We can therefore safely apply our
model to the full-band light curve (as the soft emission only provides a constant offset).

\begin{figure}
  \centering
  \includegraphics[width=\columnwidth]{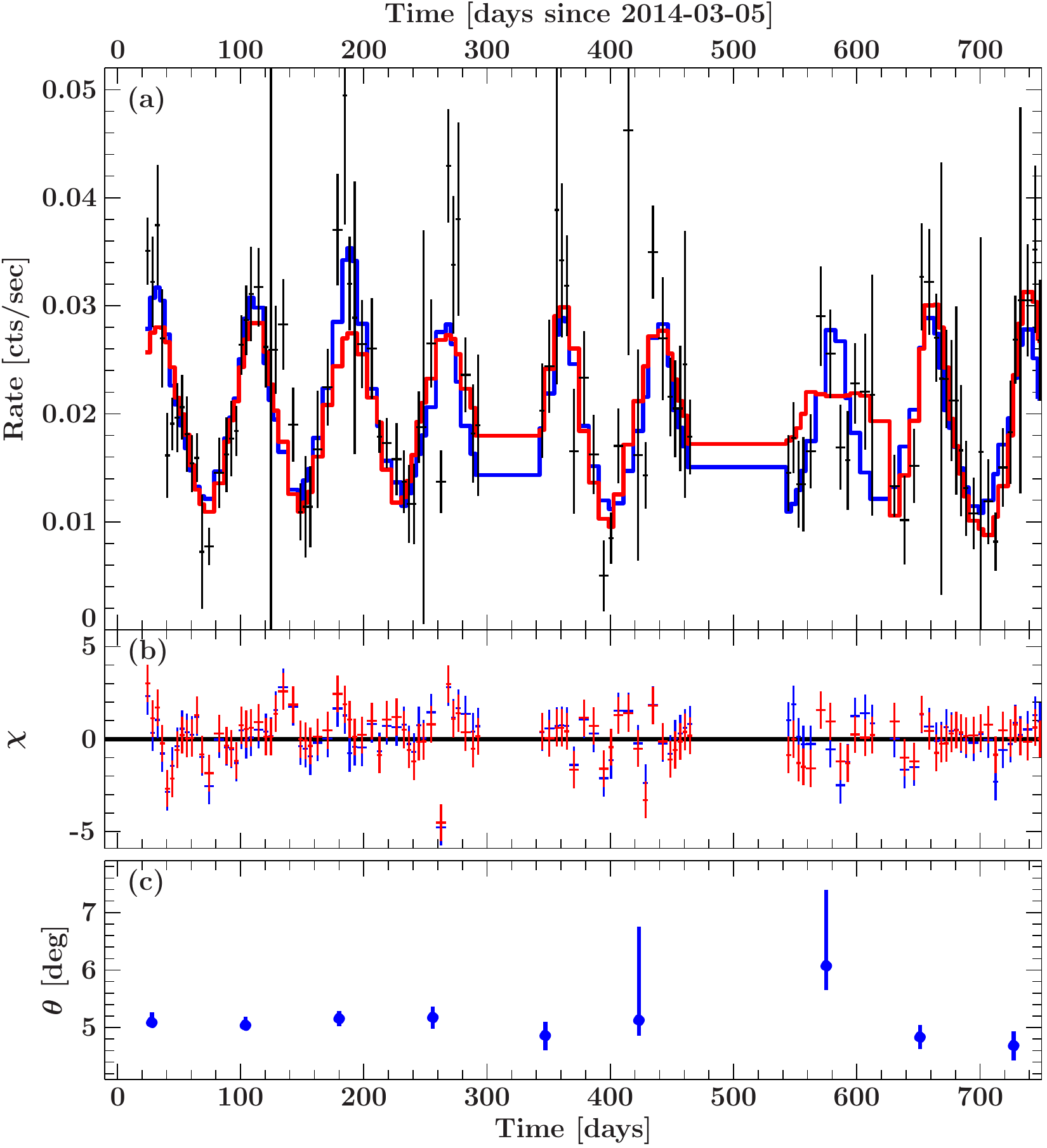}
  \caption{(a) The 0.3-10~keV \textsl{Swift} XRT light curve of NGC~5907 ULX-1,
    binned to 4 days. The model with a fixed opening angle is shown in
    red (model~1), while the model with variable opening angle is
    shown in blue (model~2). (b) Residuals of the model fits depicted
    in the same colours. (c) The half opening angles $\theta$ of the
    best fit model with a fixed intrinsic luminosity (blue
    curve in (a), model~2). In most cases the half opening angle is
    around 5 degrees. }
  \label{fig:data}
\end{figure}

The light curve with the best fit model is shown in
Fig.~\ref{fig:data}. In order to obtain reasonable results we had to
allow a shift in phase between the different precession
periods. Namely the 5th and the 6th peak are shifted by a
$\Delta\mathrm{phase}\sim 0.2$ (i.e. $\sim$14 days) with respect to
the preceding peaks (this was also visible in the light curve fitting
of Walton et al. 2016). Afterwards, the remaining three peaks coincide
directly with the first peaks (see Fig.~\ref{fig:data}a). Such a
behaviour is not unreasonable given that the precession period should
be dependent on $\dot{m}$ (Middleton et al. in prep).

As discussed above, we assume that there are no long-term
changes in $\dot{m}$ and we initially assume that the $i$,
$\theta$, and the precessing angle $\Delta i$ are constant, but that
the intrinsic source luminosity can vary between individual precession
cycles. Additionally we use a fixed period of 78.1\,days for the
precession period \citep{Walton2016a}.

Fits to the light curve show that the flux variation between different
precession cycles are all within $\pm20$\% of each other and so we
assume that the variation of $\dot{m}$ does not influence the
geometry, but only the normalisation of the peaks (i.e. the intrinsic
luminosity of the object changes). Therefore in our initial fit
(model~1) we assume the geometry to be stable and obtain a
global half opening angle $\theta$. As shown in Fig.~\ref{fig:data},
this model is able to describe the data with a goodness-of-fit of
$\chi^2 / \mathrm{d.o.f.} = 156.2/94 = 1.66$. We find an inclination
of $i = 6.40^{+0.22}_{-0.02}$ degrees and a precession angle of
$\Delta i = 7.30^{+0.13}_{-0.15}$ degrees. The half opening angle is
determined to be $\theta = 6.70^{+2.31}_{-0.11}$ degrees. The velocity
of the outflow could not be well constrained. The boosting factor
determined from this fit, assuming a representative outflow velocity of
$v=0.2\,c$ is $\mathcal{B} \approx 58$.

In a second, complementary fit (model~2), we assume that the intrinsic
luminosity is constant. Therefore the change in shape and
normalisation of the peaks will solely be described by a change in the
opening angle, i.e., the parameter $\theta$. Using these assumptions
we can find a slightly better fit $\chi^2 / \mathrm{d.o.f.} = 149.0/94
= 1.58$). In this case we find a smaller inclination of $i =
5.02^{+0.04}_{-0.07}$ with $\Delta i = 0.76^{+0.15}_{-0.18}$.

Figure~\ref{fig:data}c shows how the opening angle varies over the
different peaks for the model with a fixed (intrinsic) normalisation
(model~2). Generally, opening angles are found mainly around
$10^\circ$ (i.e., half opening angles around
$\theta=5^\circ$) with the opening angle remaining quite stable
between the different precessing periods. In the case of a slightly
smaller opening angle, we derive a large boosting factor of
$\mathcal{B} \approx 87$ for this model, again assuming a
representative outflow velocity of $v=0.2\,$c.

From Fig.~\ref{fig:data}a it can be seen that both models are able to
describe the data well and predict a similar shape for the peaks. The
relatively poor reduced $\chi^2$ values are mainly driven by a few
outlier data points, which clearly do not follow the general trend of
each peak. Therefore we conclude that the simple geometrical light
curve model is able to reproduce the overall shape of the light
curve well.

These results suggest that NGC~5907 ULX-1 is viewed almost face-on and
explains the hard X-ray spectrum \citep[see][]{Middleton2015b}. Due to
the extremely large accretion rate, it might be that the emission
comes only from a very small cone with an opening angle of roughly 10
degrees. With our estimated boosting factor ($\mathcal{B}=60$--$90$),
the observed luminosity of $\lsph \approx 8.5 \times 10^{40}
\,\mathrm{erg}/\mathrm{s}$ \citep[for a distance of 17.1\,Mpc,
  see][for more details]{Fuerst2016a} is reduced significantly to an
intrinsically emitted luminosity of $L = 1.0\mathrm{-}1.5 \times
10^{39}\,\mathrm{erg}/\mathrm{s}$. Super-critical discs have
luminosities which follow $L_{Edd}\left(1+ln(\dot{m})\right)$
\citep{Shakura1973a} implying mass accretion rates into the disc of
$\lesssim 100$-$1000$ times the Eddington rate ($\sim
1\times10^{-6}$--$1\times10^{-5} M_{\odot}/yr$ assuming a radiative
efficiency of 0.1) consistent with thermal timescale mass transfer
\citep[e.g.][]{King1999a,King1999b}.

% ================= %
\section{Summary and Conclusions}
\label{sec:summary-conclusions}
% ================= %

In this paper we have presented a simple model for scattering in
conical outflows, which are believed to exist in sources with a
super-critical accretion rate. Importantly, our model includes a
physical scattering process and so is more accurate than an analytical
approach. For radiation emitted close to the black hole, our model
predicts strong geometrical beaming of the radiation when viewing into
the cone. This beaming can lead to an effective flux boost of up to a
factor of 100 and, in cases where the conical outflow is moving at
observed speeds \citep[$\ge 0.2-0.4\,$c][]{Middleton2014a,Pinto2016a}
this value can be far higher. In order to reach such large boosting
factors, opening angles of 10 degrees and smaller are necessary. We
note, however, that for more moderate opening angles of up to 30
degrees boosting factors of up to 10 can still be achieved, hinting
that also the larger class of ULX sources observed around $\lsph = 1
\times 10^{39}\,\mathrm{erg}/\mathrm{s}$ could at least partly be
explained by the proposed geometrical beaming model \citep[although
  the majority are likely to be associated with the tail of the normal
  X-ray binary population:][]{Swartz2011a,Soria2012a,Middleton2013a}.

Observational proof that stellar compact objects are able to produce
such luminosities has recently tightened through the discovery that the ULXs M82 X-2 (Bachetti et al. 2013), NGC~7793~P13
\citep{Fuerst2016b,Israel2016b} and NGC~5907 ULX-1 \citep{Israel2016a} are
pulsars accreting at more than $100\,L_\mathrm{Edd}$. We conclude that
the model we presented is easily able to explain the observed flux
boost for these sources. Contrary to the statement of
\citep{Fuerst2016a} that the observed smooth pulse profiles in such
objects \citep[see also][]{Fuerst2016b,Bachetti2014a} is at odds with
such a narrow cone, we suggest that the different lengths of the
photon trajectory due to the multiple scatterings {\it must} smoothen
the pulse profile of the intrinsic source.

The {\sc xspec} model \texttt{ulxlc} constructed from the simulations
was successfully fitted to the long-term light curve of the ULX pulsar
NGC~5907 ULX-1. Within our model the system is found to be viewed
almost face-on with only a small precession angle needed to explain
the large quasi-periodic flux changes. We employed two complimentary
model assumptions and show that the observed difference in the peaks
of the light curve can be solely explained by intrinsic source
variability or a change in the opening angle. We note that clearly
those must be linked to some degree as the opening angle is driven by
changes in mass accretion rate. The opening angle determined from the
fits range from $10^\circ\mathrm{-}13^\circ$, resulting in a boosting
factor of roughly 60--90 and mass accretion rates $\lesssim$ 100-1000$
\times$ Eddington.

The obtained boosting factor and almost face-on geometry now strongly
support the interpretation of NGC~5907 ULX-1 with a low surface
magnetic field. Following \citet[Eq. 14]{Fuerst2016a} and
\citet{King2016a}, we can estimate the magnetospheric radius $\approx
120\,R_\mathrm{g},$, by using their spectral fitting results and
$\mathrm{cos}(i) \approx 1$. Employing the obtained intrinsic
luminosities, we derive a quite low magnetic field strength of $B
\approx 2 \times10^{10}\,$G in this case. The above results use a
pulsed fraction of 20\% \citep[see][]{Israel2016a}, which is
comparably low for a high-mass X-ray binary
\citep[see][]{Bildsten1997a}, but supporting our model where the hard
X-ray radiation is produced at the inner regions close to the neutron
star \citep[see also][]{Kluzniak2015a}.

Our simple modelling of the conical surface requires a smooth and
constant outflow, which is not expected in real RMHD flows which are
subject to instabilities \citep{Jiang2014a,Sadowski2014a} although the
resulting stratification is expected to be small scale
\citep{Takeuchi2013a} and can therefore be ignored to first
order. However, a more dedicated analysis of light curves, the
connection to the mass accretion rate, and the expected spectral
changes due to the scattering in the cone are necessary in order to
further strengthen our findings.

\emph{Acknowledgements.} TD acknowledges funding by the Deutsches
Zentrum f\"ur Luft- und Raumfahrt contract 50\,QR\,1402. MJM
acknowledges funding from STFC via an Ernest Rutherford advanced
grant. We thank Matthias K\"uhnel for valuable discussions and John
E.~Davis for the development of the \textsc{SLxfig} module used to
prepare the figures in this letter. This research has made use of ISIS
functions provided by ECAP/Remeis observatory and MIT
(http://www.sternwarte.uni-erlangen.de/isis/).

\bibliographystyle{mn2e_williams} 
\bibliography{mnemonic,mn_abbrv,local}

\appendix
\label{lastpage}

\end{document}